\documentclass{article}
\usepackage{amsmath}
\usepackage{graphicx} 
\usepackage{natbib}
\usepackage{xcolor}
\usepackage{authblk}

\title{A Review of Statistical Methods for Handling Nonignorable Missing Data using Instrument Approach}
\author[1]{Yujie Zhao}
\affil[1]{School of Industrial and Systems Engineering, Georgia Institute of Technology, Atlanta, GA, USA}

\newcommand{\diff}{\mathop{}\!\mathrm{d}}

\begin{document}
\date{}

\maketitle

\begin{abstract}
Nonignorable missing data, where the probability of missingness depends on unobserved values, presents a significant challenge in statistical analysis. Traditional methods often rely on strong parametric assumptions that are difficult to verify and may lead to biased estimates if misspecified. Recent advances have introduced the concept of a \textit{nonresponse instrument} or \textit{shadow variable} as a powerful tool to enhance model identifiability without requiring full parametric specification. This paper provides a comprehensive review of statistical methods that leverage instrumental variables to address nonignorable missingness, focusing on two predominant semiparametric frameworks: one with a parametric data model and a nonparametric propensity model, and the other with a parametric propensity model and a nonparametric data model. We discuss key developments, methodological insights, and remaining challenges in this rapidly evolving field.

    \textbf{key words:} Missing data, Nonignorable nonresponse, Instrumental variable, Shadow variable, Model identifiability.
\end{abstract}

\section{Introduction}
Missing data is frequently observed in numerous fields, including sample surveys, medical research, economics and finance, and the social sciences. 
While technology advancements have reduced the cost of data collection, they have not eliminated the impact of missing data. 
In some cases, the proportion of missing values in a dataset can be as high as 50\% to 80\%.
The most naive method for handling missing data is to discard samples that contain missing values and analyze based on the remaining complete data, which is known as the \textit{complete case} (CC) analysis. 
However, due to the diverse reasons for data absence and the variability in the proportion of missingness, CC analysis can often lead to significant bias or a substantial loss of efficiency.
In many critical sectors that affect national welfare and people's livelihoods — such as population and economic censuses, the development of new pharmaceuticals, financial risk forecasting and the guidance of public opinion — erroneous or inefficient analysis can have severe consequences. 
Therefore, it is always a major concern how to conduct statistical inference in the presence of missing data, and also a key research focus within the field of statistics.

\subsection{Missing Data Mechanism}
The most commonly used methods for analyzing missing data typically depend on the missing data mechanism, which refers to the reason the data are missing. There are three common types of missing data mechanisms.
\begin{itemize}
    \item When the missingness of data does not depend on any variables in the sample, the mechanism is known as \textit{missing completely at random} (MCAR). For example, data might be MCAR if a researcher accidentally misplaces a lab sample or if a survey respondent unintentionally forgets to answer a question.
    
    \item When the missingness of data depends only on observed variables and not on the unobserved variables themselves, the mechanism is called \textit{missing at random} (MAR). For instance, in a survey involving personal income, the absence of income data might not be related to the income level itself but could be associated with the respondent's education level. In other words, individuals with higher education may be more reluctant to disclose their specific income. The resulting data loss in this scenario is considered MAR.
    
    \item When the missingness of data is related to the unobserved values, the mechanism is termed \textit{not missing at random} (NMAR), also known as nonignorable missingness. As \cite{lipsitz1999non} pointed out, the MAR assumption often does not hold in many situations, as the probability of data being missing can still depend on the missing values even after conditioning on all other observed data. For example, individuals who are overweight might choose to conceal their true weight, and those who have engaged in dishonest or illegal activities may be more likely to evade related questions. 
\end{itemize}

It is important to note that MCAR and MAR are collectively referred to as \textit{ignorable missingness}. Over many years of development, methods for handling ignorable missing data have been thoroughly studied; see, for example, \cite{little2002statistical}, \cite{molenberghs2007missing}, \cite{kimandshao2013}, and the references therein.

For nonignorable missingness, it is more frequently observed in practice. Because the probability of missingness depends on the missing data, parameter identification and estimation under nonignorable missingness are exceptionally challenging problems. Applying complete case (CC) analysis to such data is not only inefficient but, more critically, leads directly to biased estimates and erroneous conclusions. Over the past decade, the problem of nonignorable missing data has become a research frontier in the field of missing data analysis and in statistics more broadly.

\subsection{Instrument Variable in Nonignorable Missing Data Modeling}
The problem of nonignorable missing data typically involves two models, the original data model and missing data propensity model. Maximum likelihood methods have been well developed when both models are parametric \citep{greenlees1982imputation,baker1988regression}. \cite{robins1997toward} showed that, in
order to identify all unknown parameters, either the propensity function or
the original data distribution must have a parametric component. But even when a parametric component exists, the parametric model is identifiable only under some assumptions and the identifiability issue is still difficult to handle. In recent years, a statistical tool called ``nonresponse instrument" or ``shadow variable" \citep{wang2014instrumental, zhao2015semiparametric, miaoandTchetgen2016} has been developed to address the model identification issue. Specifically, the nonresponse instrument is a useful covariate vector that can be excluded from the nonresponse propensity but are still useful covariates even when other covariates are conditioned, and it can help us to identify unknown parameters under some conditions \citep{wang2014instrumental, zhao2015semiparametric}.

\section{Review of Existing Methods to Handle Nonignorable Missing Data}

Before reviewing the relevant literature, we will introduce some notation. Let $Y$ be the response variable and $X$ be the vector of covariates. 
We denote $p(\cdot|\cdot)$ or $p(\cdot)$ as the conditional or unconditional probability density function with respect to some measure (discrete, continuous, or mixed). 
Unless otherwise specified, most of the literature mentioned in this section, as well as the primary focus of this paper, concerns the case where the response variable $Y$ has missing values. 
Let $R$ be the missing data indicator variable, where $R=1$ if the data for $Y$ is observed and $R=0$ if it is missing.

The two models involved in the problem of nonignorable missing data include: the data model, $p(Y|X)$, and the missing data propensity model, $P(R=1|Y,X)$. 
Because fully parametric models are sensitive to their underlying assumptions, most recent research involving instrumental variable methods has focused on semiparametric approaches. 
In these approaches, one of the two models — either $p(Y|X)$ or $P(R=1|Y,X)$ — is assumed to have a parametric form, while the other is assumed to be nonparametric.
Consequently, much of the research on instrumental variable methods has proceeded along two main paths:
\begin{itemize}
    \item Path 1: Assumes the data model $p(Y|X)$ is parametric, while the propensity model $P(R=1|Y,X)$ is nonparametric.
    \item Path 2: Assumes the propensity model $P(R=1|Y,X)$ is parametric, while making no assumptions about the data model $p(Y|X)$ itself.
\end{itemize}
These two paths represent the primary research directions considered in this paper. 
There have been several landmark studies along both of these paths.

\subsection{Path 1}

Under Path 1, the most common approach for addressing the problem of nonignorable missing data is the pseudo-likelihood method. 
\cite{tang2003analysis} were the first to propose using a pseudo-likelihood method to solve the parameter estimation problem under this path. 
Although their paper did not explicitly introduce the concept of an "instrumental variable," it crucially provided sufficient conditions for parameter identifiability under the pseudo-likelihood framework, thereby laying the foundation for subsequent instrumental variable-based pseudo-likelihood methods.

Under the assumption that $p(Y|X)$ has the parametric form $p(Y|X;\theta)$, where $\theta$ is the unknown parameter to be estimated, \cite{tang2003analysis} further assumed that
\begin{equation}
P(R=1|Y,X)=P(R=1|Y), \label{propensity0},
\end{equation}
that is, conditional on the response variable $Y$, the probability of $Y$ being missing is independent of the entire covariate vector $X$. \cite{tang2003analysis} first proved that under assumption \eqref{propensity0}, the parameter $\theta$ is identifiable. 
Second, by using \eqref{propensity0} and Bayes' theorem, we can obtain
\begin{equation}\label{identity0}
p(X|Y,R=1)=p(X|Y)=
\frac{p(Y|X;\theta)p(X)}{\int p(Y|x;\theta)p(x)  \diff x}.
\end{equation}
Let $\left\lbrace (y_i,x_i,r_i): i=1,..., N\right\rbrace$ be $N$ independent and identically distributed (i.i.d.) samples observed from the joint distribution of $(Y,X,R)$. According to equation \eqref{identity0}, if $p(X)$ is known, \cite{tang2003analysis} proposed that $\theta$ can be estimated by maximizing the following likelihood function:
\begin{equation}\label{pseudo0}
\prod_{i\leq N,\, r_i=1} \frac{p(y_i|x_i;\theta)p(x_i)}{\int p(y_i|x;\theta)p(x)\diff x}
\propto  \prod_{i\leq N,\, r_i=1} \frac{p(y_i|x_i;\theta)}{\int p(y_i|x;\theta)p(x)\diff x}.
\end{equation}	
In most cases, $p(X)$ in equation \eqref{pseudo0} is unknown and must be replaced by its estimate. Consequently, the updated expression \eqref{pseudo0} is not a true likelihood function. For this reason, the method is referred to as a pseudo-likelihood method.

However, in practical applications, the probability of $Y$ being missing often depends on the covariates $X$ even when conditioned on the response variable $Y$. 
Therefore, the assumption made by \cite{tang2003analysis} may not hold. 
Furthermore, \cite{wang2014instrumental} proved that if the missingness probability depends on every variable in the covariate vector $X$, then the parameter is non-identifiable.
Based on these considerations, \cite{zhao2015semiparametric} adopted a compromise assumption: they assumed that $X$ can be partitioned into two parts, $X=(U,Z)$, and that the propensity model $P(R=1|Y,X)$ satisfies
\begin{equation}\label{pseudo01}
P(R=1|Y,X)=P(R=1|Y,U),
\end{equation}
In other words, whether $Y$ is missing depends only on a subset of variables $U$ within $X$, but is independent of $Z$ (the instrumental variable). 
Essentially, then, under the assumption of \cite{tang2003analysis}, the entire covariate vector $X$ serves as the instrumental variable.
Similarly, let $\left\lbrace (y_i,u_i,z_i, r_i): i=1,..., N\right\rbrace$ be $N$ independent and identically distributed (i.i.d.) samples observed from the joint distribution of $(Y,U,Z,R)$. Under the premise that $p(U,Z)$ is either known or can be estimated, \cite{zhao2015semiparametric} based on
$$
p(Z|Y,U,R=1)=p(Z|Y,U)=
\frac{p(Y|U,Z;\theta)p(U,Z)}{\int p(Y|U,z;\theta)p(U,z)\diff z}, 
$$
to get the new pseudo-likelihood function of
\begin{equation}\label{l0}
\prod_{i\leq N,\, r_i=1} \frac{p(y_i|u_i,z_i;\theta)p(u_i, z_i)}{\int p(y_i|u_i,z;\theta)p(u_i, z)\diff z}
\propto  \prod_{i\leq N,\, r_i=1} \frac{p(y_i|u_i,z_i;\theta)}{\int p(y_i|u_i,z;\theta)p(u_i, z)\diff z}.
\end{equation}	
By maximizing the likelihood function in \eqref{l0}, we get the pseudo-likelihood estimate of $\theta$.
Finally, \cite{zhao2015semiparametric} applied this method to generalized linear models, while \cite{shao2013estimation} further extended it to panel data. \cite{fang2016regression} and \cite{fang2018imputation} combined the pseudo-likelihood and imputation methods to handle cases with missing covariates. \cite{chen2018semiparametric} studied parameter estimation and statistical inference of estimating equations when covariates contain nonignorable missing data. \cite{chen2018pseudo} proposed a modification of the pseudo-likelihood method based on nonparametric kernel regression and sufficient dimension reduction (SDR), which avoids making any parametric model assumptions on the joint distribution of high-dimensional covariates and achieves dimension reduction under nonignorable missing data.

Although the pseudo-likelihood method does not make any parametric assumptions about the propensity model $P(R=1|Y,X)$, it still requires the data model $p(Y|X)$ to have a fully parametric form. To further mitigate the sensitivity of the pseudo-likelihood method to these parametric assumptions, \cite{Fang2016} proposed a penalized validation criterion (PVC) to address the problem of selecting the data model $p(Y|X)$.
Assume there exists a pre-specified set of candidate parametric models for $p(Y|X)$ is
\[
\mathcal{M}=\{ M_j :p_j(Y|X;\theta^{(j)}), j=1,\dots,J\},
\]
where $\theta^{(j)}$ is the unknown parameter vector for model $M_j$, and at least one model in $\mathcal{M}$ is the correctly specified model for $p(Y|X)$. 
The objective of \cite{Fang2016} is to select a model from $\mathcal{M}$ that is not only correct but also the most efficient.
The core idea of their paper is to compare two estimators of the conditional distribution $F_1(z)=P(Z\leq z|R=1)$, where $z$ is a possible value of the instrumental variable $Z$. 
The first estimator is the empirical cumulative distribution function, 
$
\tilde{F}1(z) = \frac{\sum{i=1}^N I(z_i \leq z, r_i=1)}{\sum_{i=1}^N r_i}
$, where $I(\cdot)$ is the indicator function. 
The second is an estimator constructed based on the pseudo-likelihood method, denoted as $\hat{F}_1(z)$.
It should be noted that $\tilde{F}_1(z)$ is always a consistent estimator of the true distribution $F_1(z)$. Based on the theoretical properties of the pseudo-likelihood method, $\hat{F}_1(z)$ is a consistent estimator only if $p(Y|X;\theta^{(j)})$ is the correctly specified model for $p(Y|X)$. Therefore, \cite{Fang2016} proposed the following validation criterion (VC) to measure the distance between $\tilde{F}_1(z)$ and $\hat{F}_1(z)$:
\begin{equation*}
\text{VC} =\frac{1}{N} \sum_{i=1}^{N} \big|\tilde F_{1}(z_{i})- \hat F_1(z_{i})\big|.
\end{equation*}
A smaller VC value indicates that the corresponding model is more likely to be the correct one.
To avoid the problem of model overfitting, \cite{Fang2016} further proposed the following penalized validation criterion (PVC):
\begin{equation*}
\text{PVC}= \text{VC} +\lambda \log d,
\end{equation*}
where $\lambda_N$ is a penalty coefficient and $d_j$ is the dimension of the unknown parameter vector $\theta^{(j)}$ in model $M_j$. 
The model corresponding to the smallest $PVC$ value is selected as the most efficient data model.

Furthermore, when the data model is correctly specified, \cite{chen2021instrument} proposed a pseudo-likelihood approach to search for an instrument from a given set of covariates. Afterwards, by combining the approaches of \cite{Fang2016} and \cite{chen2021instrument}, we can achieve simultaneous selection of instrumental variables, covariates, and data parameter models, see \citet{chen2025instrumentvariablemodelselection} for more details.

\subsection{Path 2}
Under Path 2, since the data model $p(Y|X)$ is nonparametric, estimating the parameters of the propensity model $P(R=1|Y,X)$ alone is often not of sufficient practical interest. 
Therefore, most research in this area that uses instrumental variables also focuses on estimating other population parameters, such as the mean of $Y$. 
Due to the presence of nonignorable missing data, even estimating a quantity as simple as the mean of $Y$ is often a challenging task.

The earliest research along this path using instrumental variables was conducted by \cite{wang2014instrumental}. 
This paper first discusses the issue of parameter identifiability for instrumental variable methods and proves that when no instrumental variable $Z$ satisfying condition \eqref{pseudo01} exists, parameters in the model may be non-identifiable. 
It is important to note that this problem is not confined to semiparametric methods; the issue of parameter non-identifiability can persist even when both the data model $p(Y|X)$ and the propensity model $P(R=1|Y,X)$ are assumed to have fully parametric forms.
Next, it is assumed that the propensity model has the following parametric form:
\[
P(R=1|Y,U)=\Psi(\alpha+\beta U +\gamma Y),
\]
where $\Psi(\cdot)$ is a known function and $\vartheta=(\alpha,\beta,\gamma)$ is the vector of unknown parameters.
\cite{wang2014instrumental} used the generalized method of moments (GMM) to estimate the unknown parameters in the propensity model, and subsequently used these estimates to estimate the population mean of the response variable $Y$. 
To facilitate the description of the core idea behind this estimation method, we will assume here that all variables in $Z$ are continuous. The case where $Z$ contains discrete variables can be found in \cite{wang2014instrumental}. 
First, the following equation is constructed:
\begin{equation}
    g(Y,X,R,\vartheta)=\left(\begin{array}{c}1\\U\\Z\end{array}\right)\left\{\frac{R}{\Psi(\alpha+\beta U +\gamma Y)}-1\right\},
    \label{g50}
\end{equation}
which satisfies the condition $E[g(Y,X,R,\vartheta)]=0$.
Since the dimension of $Z$ may be greater than the dimension of $Y$, which implies that the dimension of $g$ in \eqref{g50} is greater than the dimension of $\vartheta$, we can obtain an estimate of $\vartheta$ as $\hat \vartheta=(\hat \alpha,\hat \beta,\hat \gamma)$, by directly applying the generalized method of moments (GMM) to $g$.
Next, using the inverse probability weighting (IPW) method, the estimator for the population mean of $Y$ is given by:
\[
\frac 1 N\sum_{i=1}^N \frac{r_iy_i}{\Psi(\hat \alpha+\hat \beta u_i +\hat \gamma y_i)}.
\]

It is important to note that the method of \cite{wang2014instrumental} assumes a fully parametric model for the propensity score. 
To mitigate the impact of potential model misspecification, \cite{shao2016semiparametric}, following the approach of \cite{kim2011semiparametric}, further assumed that the propensity model has a semiparametric exponential tilting form:
\begin{equation}\label{propensitywang}
P(R=1|Y,U)=[1+\exp\{h(U)+\gamma Y \}]^{-1},
\end{equation}
where $h(\cdot)$ is an unknown function and $\gamma$ is an unknown parameter. 
Under the assumption that condition \eqref{propensitywang} holds, \cite{shao2016semiparametric} extended the method of \cite{wang2014instrumental} to estimate the population parameters.

Furthermore, \cite{wangshaofang2018} addressed the problem of simultaneously selecting the instrumental variable and the propensity model under Path 2. \cite{chen2019semiparametric} studied parameter estimation and statistical inference of estimating equations in the presence of nonignorable missing data for the response variable.
In a setting where both the instrumental variable $Z$ and the propensity model $P(R=1|Y,U)$ are not fixed,  \cite{wangshaofang2018}, following the approach of \cite{Fang2016}, constructed two estimators for the distribution of the covariate vector $X$. 
They then defined a similar penalized validation criterion to select the correct and most efficient combination of $Z$ and $P(R=1|Y,U)$. 

\subsection{Other Paths}
In addition to the two common paths described above, \cite{miaoandTchetgen2016} approached the problem of reducing sensitivity to parametric model assumptions from a different angle. 
They proposed using an instrumental variable to obtain doubly robust estimators for the mean of $Y$ in the presence of a missing response variable.
This doubly robust estimator is constructed by positing three models:
\begin{itemize}
    \item $P(R=1|Y=0,U)$
    \item $p(Z,Y|R=1,U)$
    \item A log-odds ratio model that characterizes the relationship between the response variable $Y$ and the missingness mechanism.
\end{itemize}
\cite{miaoandTchetgen2016} proved that if the log-odds ratio model is correctly specified, then only one of the other two models needs to be correct to ensure the consistency of the doubly robust estimator.

Finally, due to the rapid development of information technology in recent years, high-dimensional data has become a common phenomenon. 
Statistical analysis in the presence of both high-dimensional and missing data has consistently been a focal point in the field of statistical research.
When high-dimensional data contains ignorable missing values, a body of literature has approached the problem from the perspective of sufficient dimension reduction (SDR), including studies by \cite{hu2010semiparametric}, \cite{ding2011fusion}, \cite{deng2017dimension}, and \cite{li2017mean}. 
Another line of research addresses the variable selection problem through feature screening, as seen in the work of \cite{lai2017model}, \cite{wang2018make}, \cite{tang2019feature}, \cite{zambom2019sure} and \cite{ni2020feature}
When high-dimensional data contains nonignorable missing values, some studies have also explored the problem from the perspective of dimension reduction or feature screening; see, for example, \cite{wang2019} and \cite{zhang2019}.

\section{Conclusion and Discussion}

\begin{figure}[htbp]
	\centering
	\caption{Nonignorable missing data literature review\label{fig1}}
    \includegraphics[width=0.98\textwidth]{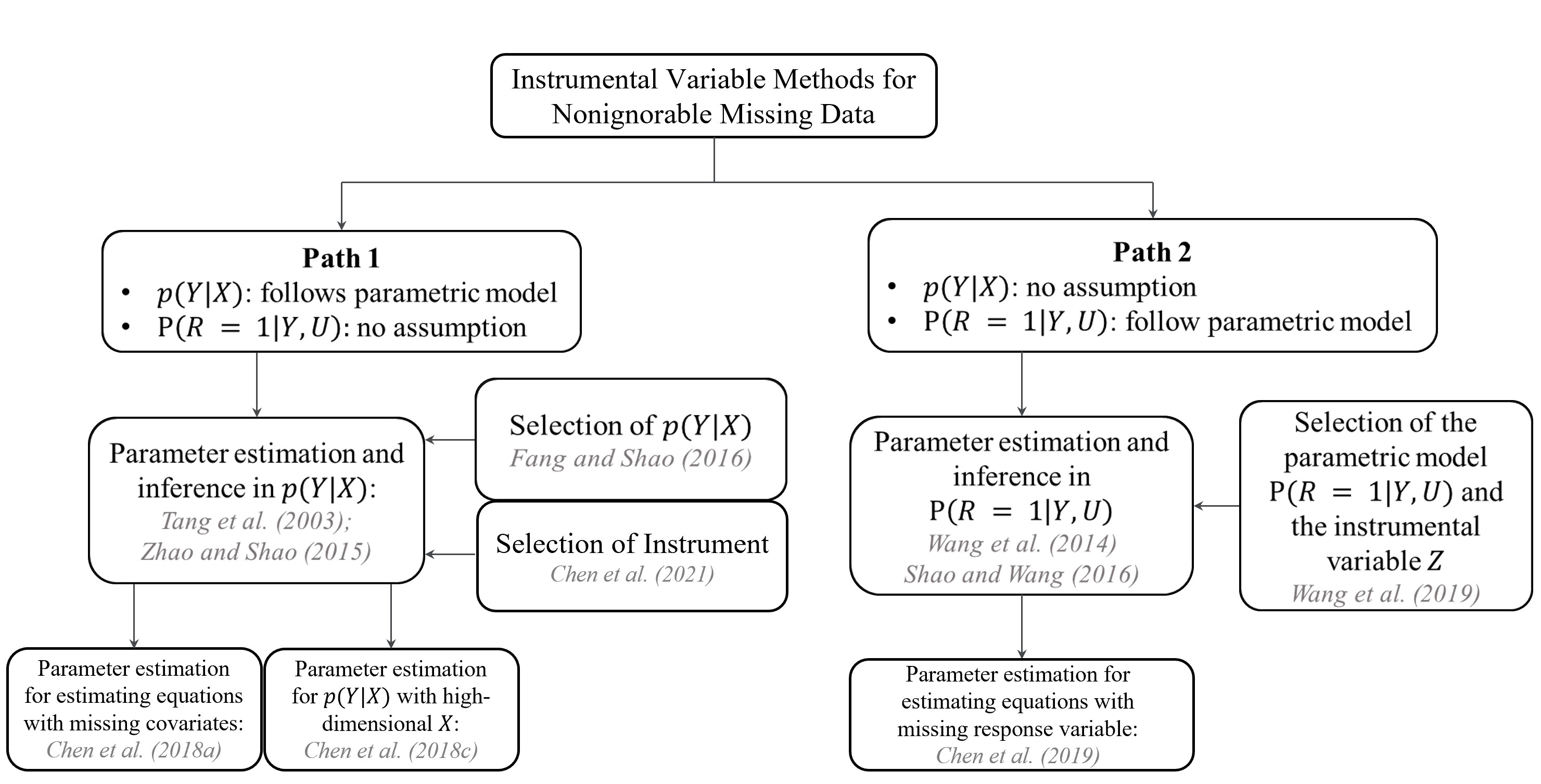}
\end{figure}

This review has surveyed a range of methodological advances in handling nonignorable missing data through the use of instrumental variables, as illustrated in Figure \ref{fig1}. We have focused on two main semiparametric pathways: one that assumes a parametric data model with a nonparametric propensity function, and another that assumes a parametric propensity model with a nonparametric data distribution. Each approach offers distinct advantages in terms of identifiability and estimation, and both have spurred significant research in areas such as pseudo-likelihood estimation, model selection, and doubly robust methods.

Despite these advances, several important challenges remain. First, while some methods have been extended to high-dimensional settings, most are not yet suited to scenarios where the number of covariates exceeds the sample size. Second, complex data structures with nonignorable missingness require further methodological development, such as longitudinal or functional data \citep{tseng2016longitudinal, chen2018functional, lin2023flexible}, or survival data \citep{Zhang_Ling_Zhang_2024}. Third, although model selection and doubly robust estimators help mitigate specification bias, their performance still depends on the correctness of underlying assumptions. The doubly robust framework, in particular, offers promising protection against model misspecification, but its practical implementation warrants further study.

Future research should focus on extending instrumental variable methods to ultrahigh-dimensional settings, adapting them for complex data types, and refining robust estimation strategies. Continued exploration in these areas will be essential for improving the analysis of nonignorable missing data across scientific disciplines. As these methodologies mature and become more robust, they hold significant potential for broad application across diverse fields. For instance, in medical research, they could improve causal inference from electronic health records or clinical trials with selective documentation \citep{little2012prevention, ratitch2013missing}. For registry data, these approaches could address systematic non-response in population-level databases \citep{wei2018u, chen2021natural, gottlieb2025differences, gottlieb2025work, openshaw2025effect}. In social sciences, they could help correct for non-random attrition in longitudinal surveys \citep{faust2025firearm, nakatsuka2025all}. The development of more general and computationally efficient frameworks will enable researchers across these domains to address the pervasive challenge of nonignorable missing data with greater rigor and reliability.

\bibliographystyle{apalike} 
\bibliography{reference}
\end{document}